\documentstyle[prd,aps,preprint,psfig]{revtex}

\begin{document}
\tighten
\title{$R$-Invariant Topological Inflation}
\author{Izawa K.-I.} \address{Department of Physics,
  University of Tokyo, Tokyo 113-0033, Japan}
\author{M. Kawasaki}
\address{Institute for Cosmic Ray Research, University of Tokyo,
  Tanashi 188-8502, Japan}
\author{T. Yanagida} \address{Department of Physics and RESCEU,
  University of Tokyo, Tokyo 113-0033, Japan}
\date{\today}

\maketitle
\begin{abstract}
    We propose a topological inflation model in the framework of
    supergravity with $R$ invariance. This topological inflation model
    is not only free from the initial value problem of the inflaton
    field but also gives low reheating temperature which is favored in
    supergravity since the overproduction of gravitinos is avoided.
    Furthermore, the predicted spectrum of the density fluctuations
    is generally tilted, which will be tested by future observations
    on CMB anisotropies and large scale structure of the universe.
\end{abstract}



The topological~\cite{Linde,Vilenkin} and chaotic~\cite{Linde-Book}
inflation models are very attractive among various inflation models
considered so far, since they do not suffer from the initial value
problem~\cite{Linde-Book} for an inflaton field $\varphi$. That is,
the inflationary universe is a natural consequence of the system. The
chaotic inflation model assumes large initial values $\varphi_0 \gg
1$, where we set the gravitational scale $M = 2.4 \times 10^{18}$GeV
to unity.  This is, however, unlikely realized in the framework of
supergravity~\cite{Lyth}. Thus, we are led to consider the topological
inflation model.

Naive models of topological inflation in supergravity
(see Ref.\cite{Izawa})
tend to yield anti-de Sitter universe at the end of inflation.
To go around this problem, in this paper, we propose
a topological inflation model with the $R$-invariant vacuum
which ensures vanishing cosmological constant
at the end of inflation.\footnote{
For $R$-invariant models of subjects other than inflation, see
Ref.\cite{Yanagida}}

Let us consider a superpotential
\begin{equation}
    W = \Lambda^2 Z (\lambda 
        - \lambda' \phi^2) = v^2 Z (1 - g \phi^2),
\end{equation}
where $\Lambda$ denotes a scale presumably generated
dynamically\footnote{
For example, the scale may arise from a hyperquark condensation
$\langle Q{\bar Q} \rangle = \Lambda^2$ with the aid of a
superpotential considered in Ref.\cite{Hotta}.}
and $v^2 = \lambda \Lambda^2$, $g = \lambda'/\lambda$.
Here, we impose $U(1)_R \times Z_{2}$ symmetry and omit higher-order
terms for simplicity. Under the $U(1)_{R}$
we assume
\begin{equation}
    Z(\theta)  \rightarrow  e^{-2i\alpha} Z(\theta e^{i\alpha}),~~~
    \phi(\theta)  \rightarrow \phi(\theta e^{i\alpha}).
\end{equation}
We also assume that $Z$ is even and 
$\phi$ is odd under the $Z_2$.
This discrete $Z_2$ symmetry is an essential ingredient for the
topological inflation \cite{Linde,Vilenkin}.

The $R$- and $Z_2$-invariant natural K\"ahler potential is given by
\begin{equation}
    \label{new-kpot}
    K(\phi,Z) = |Z|^2 + |\phi|^2 + k_{1}|Z|^2|\phi|^2 
    -\frac{k_{2}}{4}|Z|^4
    + \cdots ,
\end{equation}
where $k_{1}$ and $k_{2}$ are constants of order one.

The potential of a scalar component of the superfields
$X(x,\theta)$ and $\phi(x,\theta)$ in supergravity is
given by
\begin{equation}
    \label{potential}
    V = e^{K} \left\{ \left(
        \frac{\partial^2K}{\partial z_{i}\partial z_{j}^{*}}
      \right)^{-1}D_{z_{i}}W D_{z_{j}^{*}}W^{*}
      - 3 |W|^{2}\right\},~~~~~~~(z_i = \phi, Z )
\end{equation}
with 
\begin{equation}
        D_{z_i}W = \frac{\partial W}{\partial z_{i}} 
        + \frac{\partial K}{\partial z_{i}}W.
        \label{DW}
\end{equation}
This potential yields an $R$-invariant vacuum 
\begin{eqnarray}
    \langle Z \rangle = 0,
    \quad
    \langle \phi \rangle = 
    \frac{1}{\sqrt{g}} \equiv \eta,
    \label{eq:eta-g}
\end{eqnarray}
in which the potential energy vanishes. Here, the scalar components of
the superfields are denoted by the same symbols as the corresponding
superfields.  The inflaton $\phi$ has a mass $m_{\phi}$ in the vacuum
with
\begin{equation}
    \label{inflaton-mass}
    m_{\phi} \sim 2|\sqrt{g}v^2|.
\end{equation}

A topological inflation~\cite{Linde,Vilenkin} may occur if the vacuum
expectation value $\langle \phi \rangle$ is of the order of the
gravitational scale.  The initial values of $\phi$ are randomly
distributed in spaces owing to the chaotic condition in the early
universe. When $\phi$ settles down to the minima of the potential
(\ref{potential}), domain walls appear since the $Z_2$ symmetry is
spontaneously broken in the vacua.  The thickness of the domain wall
is as large as the horizon size and the false vacuum energy inside the
wall causes inflation \cite{Linde,Vilenkin} if the potential is
sufficiently flat for $|\phi| \lesssim 1$.  The critical value
$\eta_{\rm cr}$ of $\langle \phi \rangle$ for which the topological
inflation occurs was investigated in Refs.\cite{Sakai,Cho,Laix}. We
adopt the result of Ref.~\cite{Sakai}, which gives
\begin{equation}
     \eta_{\rm cr} \simeq 1.7.
\end{equation}
Thus, for topological inflation to take place, $|\eta|$ should be
larger than $\eta_{\rm cr}$, which leads to the constraint on $g$:
\begin{equation}
    \label{g-v}
    \eta_{\rm cr} |\sqrt{g}| < 1.
\end{equation}

Now let us discuss the dynamics of the topological inflation. From
eq.(\ref{potential}) the potential for $|Z|$ and $|\phi| \ll 1$
is written approximately as
\begin{equation}
    \label{eff-pot}
    V \simeq v^4|1 - g\phi^{2}|^{2} + (1 - k_1)v^4 |\phi|^2 
    + k_2v^4 |Z|^2.
\end{equation}
Since the $Z$ field quickly settles down to the origin for $k_2
\gtrsim 1$, we set $Z =0$ in eq.(\ref{eff-pot}) assuming $k_2 \gtrsim
1$.  Note that the $R$ symmetry is not broken even during the
inflation.  For $g > 0$, we can identify the inflaton field
$\varphi(x)/\sqrt{2}$ with the real part of the field $\phi(x)$ since
the imaginary part of $\phi$ has a positive mass with the parameters
we use in the following analysis. We
obtain a potential for the inflaton,
\begin{equation}
    \label{eff-pot2}
    V(\varphi) \simeq v^4 - \frac{\kappa}{2}v^4\varphi^2
\end{equation}
where $\kappa \equiv 2g + k_{1} - 1$.

The slow-roll condition for the inflaton is satisfied for $0< \kappa
< 1$ and $\varphi < \varphi_f$ where $\varphi_f$ is of order one,
which provides the value of $\varphi$ at the end of inflation. The
Hubble parameter during the inflation is given by
\begin{equation}
    \label{new-hubble}
    H \simeq \frac{v^2}{\sqrt{3}}.
\end{equation}
The scale factor of the universe increases by a factor of $e^N$ when 
the inflaton $\varphi$ rolls slowly down the potential from 
$\varphi_N$ to $\varphi_f$.  The $e$-fold number $N$ is given by
\begin{eqnarray}
    N \simeq \int^{\varphi_{N}}_{\varphi_f}
      d\varphi \frac{V}{V'}
      \simeq {1 \over \kappa}\ln{\varphi_f \over \varphi_N}.
    \label{N-efold}
\end{eqnarray}

The amplitude of primordial density fluctuations $\delta \rho/\rho$ 
due to this inflation is written as 
\begin{equation}
    \label{density}
    \frac{\delta\rho}{\rho} \simeq \frac{1}{5\sqrt{3}\pi}
    \frac{V^{3/2}(\varphi_{N})}{|V'(\varphi_{N})|}
    \simeq \frac{1}{5\sqrt{3}\pi} \frac{v^{2}}{\kappa\varphi_{N}}.
\end{equation}
This  should be normalized to the
data of anisotropies of the cosmic microwave background radiation
(CMB) by the COBE satellite.  Since the $e$-fold number $N$
corresponding to the COBE~\cite{COBE} scale is about $60$, the COBE
normalization gives
\begin{equation}
    \label{COBE-norm}
    \frac{V^{3/2}(\varphi_{60})}{|V'(\varphi_{60})|}
    \simeq 5.3\times 10^{-4},
\end{equation}
Along with eq.(\ref{N-efold}), we obtain
\begin{eqnarray}
    v  \simeq  2.3 \times 10^{-2} \sqrt{\kappa \varphi_f} 
    e^{-\frac{\kappa N}{2}}
    \simeq 1.8 \times 10^{-3} - 2.5 \times 10^{-5},
\end{eqnarray}
for $\varphi_f \simeq 1$ and  $0.02 \le \kappa \le 0.2$.

The interesting point on the above density fluctuations is that it
results in the tilted spectrum whose spectrum index $n_s$ is given by
\begin{equation}
    \label{new-index}
    n_s \simeq 1 - 2 \kappa.
\end{equation}
We may expect a possible deviation from the Harrison-Zeldvich
scale-invariant spectrum $n_s = 1$. For example, $n_s \simeq 0.8$
for $\kappa = 0.1$.

\begin{figure}[htbp]
  \begin{center}
   \centerline{\psfig{figure=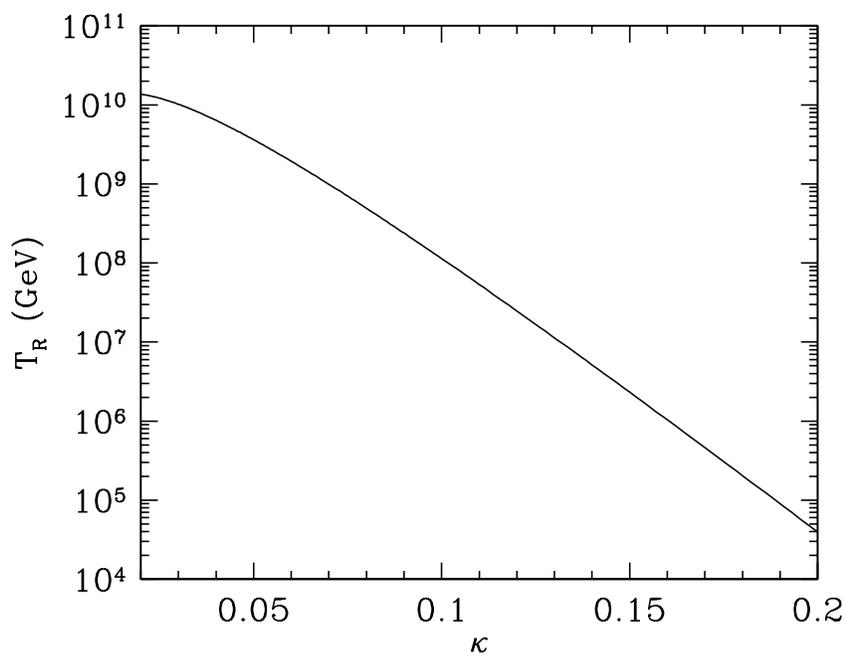,width=12cm}}
   \vspace*{-4cm}
    \caption{Reheating temperature $T_{R}$ as a function of $\kappa$. We take
    $\eta=\eta_{\rm cr}$.}
    \label{fig:rtemp}
  \end{center}
\end{figure}

After inflation ends, the inflaton $\phi$ may decay into ordinary
particles through $R$- and $Z_2$-invariant interactions with the ordinary light
fields $\psi_i$ in the K\"ahler potential:
\begin{equation}
    K(\phi,\psi_i) = \sum c_i |\phi|^2|\psi_i|^2,
\end{equation}
where $c_i$ is a coupling constant of order one and we neglect
higher-order terms for simplicity. With these
interactions the decay rate $\Gamma_{\phi}$ of the inflaton is
estimated as
\begin{equation}
    \Gamma_{\phi} \simeq \sum_i c_i^2\eta^2
    m_{\phi}^3,
\end{equation}
which yields reheating temperature $T_R$ given by
\begin{equation}
    \label{reheat-temp}
    T_R \simeq 0.46 c\eta
    m_{\phi}^{3/2},
\end{equation}
where $c \equiv \sqrt{\sum_i c_i^2}$. Using Eqs.(\ref{eq:eta-g}) and
(\ref{inflaton-mass}) we obtain
\begin{equation}
    m_{\phi} \sim 2 {v^2 \over \eta}.
\end{equation}
Thus, the reheating temperature is estimated as
\begin{equation}
    T_R 
    \sim  1.4 \times 10^{10}{\rm GeV} - 4.0 \times 10^4{\rm GeV},
\end{equation}
for $c \simeq 1$, $\eta \simeq \eta_{\rm cr}$ and $0.02 \le \kappa \le 0.2$
(Fig.~\ref{fig:rtemp}). The
predicted reheating temperature is low enough to avoid overproduction
of gravitinos~\cite{Ellis,Moroi} in a wide range of gravitino mass.

In summary, we have studied how the topological inflation takes place
in the framework of supergravity with $R$ invariance. The present
model has a couple of attractive points. First of all, since the
topological inflation naturally occurs for the chaotic condition of
the early universe, it is free from the initial value problem of the
inflaton field. In other words, the inflation is an automatic
consequence of the dynamics of the system.  The present model shares
the same merit as the new inflation model~\cite{Izawa} in that the
reheating temperature is low enough to avoid the overproduction of
gravitinos. Furthermore, the predicted spectrum of the density
fluctuations are generally tilted because the mass term induced from
the K\"ahler potential is comparable to Hubble parameter during
inflation. This tilted spectrum will be tested by future observations
on CMB anisotropies and large scale structure of the universe.

\vspace{0.5cm}
\noindent
{\bf Acknoledgement}

MK and TY are supported in part by the Grant-in-Aid, Priority Area
``Supersymmetry and Unified Theory of Elementary Particles''(\#707).

\begin{references}
\bibitem{Linde} 
  A. Linde,
  Phys. Lett. {\bf B327}, 208 (1994).
\bibitem{Vilenkin}
  A. Vilenkin,
  Phys. Rev. Lett. {\bf 72}, 3137 (1994).
\bibitem{Linde-Book} 
  For example, A.D. Linde,
  Particle Physics and Inflationary Cosmology, 
  (Harwood, Chur, Switzerland, 1990).
\bibitem{Lyth}
  For a review, D.H. Lyth and A. Riotto, hep-ph/9807278.
\bibitem{Izawa}
  Izawa K.-I. and T. Yanagida, Phys. Lett. {\bf B393}, 331 (1997); \\
  Izawa K.-I., M. Kawasaki and T. Yanagida, Phys. Lett. {\bf B411}, 249 (1997).
\bibitem{Yanagida}
  Izawa K.-I. and T. Yanagida, Prog. Theor. Phys. {\bf 97}, 913 (1997);
  {\bf 99}, 423 (1998); hep-ph/9809366.
\bibitem{Hotta}
  T. Hotta, Izawa K.-I. and T. Yanagida, Phys. Rev. {\bf D55}, 415 (1997); \\
  Izawa K.-I., Y. Nomura, K. Tobe and T. Yanagida, Phys. Rev. {\bf D56},
  2886 (1997).
\bibitem{Sakai}
  N. Sakai, H. Shinkai, T. Tachizawa and K. Maeda,
  Phys. Rev. {\bf D53}, 655 (1996); \\
  N. Sakai,
  Phys. Rev. {\bf D54}, 1548 (1996).
\bibitem{Cho}
  I. Cho and A. Vilenkin,
  Phys. Rev. {\bf D56}, 7621  (1998).
\bibitem{Laix}
  A.A. de Laix, M. Troden and T. Vachaspati
  Phys. Rev. {\bf D57}, 7186  (1998).
\bibitem{COBE}
  C.L. Bennett {\it et al.},
  Astrophys. J. {\bf 464}, L1 (1996).
\bibitem{Ellis}
  M. Yu. Khlopov and A.D. Linde,
  Phys. Lett. {\bf B138} (1984) 265; \\
  J. Ellis, E. Kim and D.V. Nanopoulos,
  Phys. Lett. {\bf B145} (1984) 181; \\
  J. Ellis, G.B. Gelmini, J.L. Lopez, D.V. Nanopoulos and S. Sarker,
  Nucl. Phys. {\bf B373} (1992) 399; \\
  M. Kawasaki and T. Moroi,
  Prog. Theor. Phys. {\bf 93} (1995) 879.
\bibitem{Moroi} 
  T. Moroi, H. Murayama and M. Yamaguchi,
  Phys. Lett. {\bf B303} (1993) 289.
\end{references}
\end{document}